\documentclass[aps,showpacs,prb,twocolumn,superscriptaddress]{revtex4}
\usepackage{amsmath}
\usepackage{graphicx}
\usepackage{booktabs}
\usepackage{multirow}
\usepackage[mediumspace,mediumqspace,Grey,squaren]{SIunits}
\usepackage{color}
\begin{document}

\title{Ultrafast Magnetoelastic Probing of Surface Acoustic Transients}

\author{J. Janu\v{s}onis}
\thanks{These authors contributed equally}
\affiliation{Zernike Institute for Advanced Materials, University
of Groningen, Groningen, The Netherlands}
\author{C.L. Chang}
\thanks{These authors contributed equally}
\affiliation{Zernike Institute for Advanced Materials, University
of Groningen, Groningen, The Netherlands}
\author{T. Jansma}
\affiliation{Zernike Institute for Advanced Materials, University
of Groningen, Groningen, The Netherlands}
\author{A. Gatilova}
\affiliation{Zernike Institute for Advanced Materials, University
of Groningen, Groningen, The Netherlands}
\author{A.M. Lomonosov}
\affiliation{IMMM CNRS 6283, Universit\'e du Maine, 72085 Le Mans
cedex, France}
\author{V. Shalagatskyi}
\affiliation{IMMM CNRS 6283, Universit\'e du Maine, 72085 Le Mans
cedex, France}
\author{V.S. Vlasov}
\affiliation{IMMM CNRS 6283, Universit\'e du Maine, 72085 Le Mans
cedex, France}
\author{V.V. Temnov}
\affiliation{IMMM CNRS 6283, Universit\'e du Maine, 72085 Le Mans
cedex, France} \affiliation{Fritz-Haber-Institut der
Max-Planck-Gesellschaft, Abteilung Physikalische Chemie,
Faradayweg 4-6, 14195 Berlin, Germany}
\author{R.I. Tobey}
\email{r.i.tobey@rug.nl} \affiliation{Zernike Institute for
Advanced Materials, University of Groningen, Groningen, The
Netherlands}

\begin{abstract}
We generate in-plane magnetoelastic waves in nickel films using
the all-optical transient grating technique.  When performed on
amorphous glass substrates, two dominant magnetoelastic
excitations can be resonantly driven by the underlying elastic
distortions, the Rayleigh Surface Acoustic Wave and the Surface
Skimming Longitudinal Wave.  An applied field, oriented in the
sample plane, selectively tunes the coupling between magnetic
precession and one of the elastic waves, thus demonstrating
selective excitation of coexisting, large amplitude magnetoelastic
waves.  Analytical calculations based on the Green's function
approach corroborate the generation of the non-equilibrium surface
acoustic transients.

\end{abstract}
\maketitle Generating elementary excitations at solid surfaces and
interfaces underscores many processes in materials and enhances
their use in modern technology.  With the increased emphasis on
new materials and enhanced functionality of existing materials,
generating and detecting multiple, competing excitations at the
surface provides opportunities to expanded implementation. As many
of these excitations are transient in nature, the use of
ultrashort optical pulses provides for the generation and
real-time monitoring of their dynamics, and allows for the
time-domain identification of their effects on the state of the
material \cite{Temnov}.

The effects of competing excitations on material properties are
exemplified in plasmonics. In research related to extraordinary
transmission of light through subwavelength
apertures\cite{Ebbesen, Lezec, Gay} a long-lasting controversy
exists over the nature of the responsible mechanism.  It is
currently understood that two competing excitations, the so-called
Composite Diffraction Evanescent Waves (CDEW) and the conventional
Surface Plasmon Polaritons (SPPs), conspire to enhance
transmission through apertures, while their contributions depend
strongly on the experimental geometry. Alternatively, new model
systems are sought in order to shed further light on extraordinary
transmission effects, and acoustic analogues have been
demonstrated\cite{Lu}.

In this Letter we report on the generation and selective detection
of two surface acoustic waves, which are analogous to CDEW and SPP
in plasmonics.  The combination of femtosecond transient grating
(TG) excitation with ultrafast time-resolved magneto-optical
spectroscopy allows for an unambiguous observation of Rayleigh
Surface Acoustic Wave (SAW) and a short-living Surface Skimming
Longitudinal Wave (SSLW), both of which couple to the
magnetization of the material.  The differences in the acoustic
properties between SAW and SSLW provides the possibility to
selectively probe the individual acoustic modes via the resonant
magneto-elastic excitation, and represents an advantage when
compared to the analogous plasmonic investigations, particularly
in regards to the selectivity in detection between the two
competing excitations.  The combination of novel experimental and
analytical theoretical tools represents a powerful playground for
the design of ultrafast magneto-optical and magneto-acoustic
devices. As the most straight-forward application, our results can
contribute to the detailed understanding of the extraordinary high
acoustic transmission through the periodically microstructured
surfaces, and by searching for novel phenomena in the
magneto-optical transmission measurements through sub-wavelength
hole arrays patterned in hybrid metal-ferromagnet multilayers. In
a broader context, the knowledge of SAW and SSLW generation in
complex nanostructures can help tailor the thermal transport
through interfaces.

\begin{figure}
\includegraphics[width=.6\columnwidth]{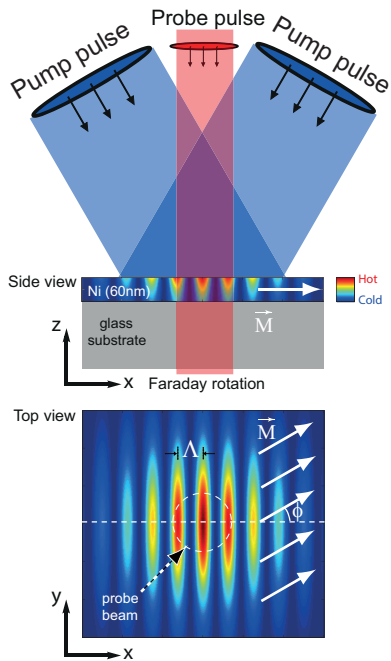}
\caption{Experimental geometry showing the thin nickel film with
glass substrate.  The transient grating is generated by two
crossed femtosecond laser pump pulses, leading to a
spatially-periodic impulsive heating of a thin nickel film and the
launching of acoustic waves along the surface of the semi-infinite
glass substrate. Grating periodicities as small as 1$~\mu$m can be
achieved.  A magnetic field can be rotated continuously in the
sample plane.  Faraday rotation of time-delayed optical probe
pulses transmitted through the sample monitors the interaction
between elastic and magnetic degrees of freedom.} \label{fig1}
\end{figure}

We recently demonstrated a novel excitation geometry for
generating magnetoelastic waves, whereby narrowband planar elastic
waves were shown to resonantly drive planar magnetization
precession using Rayleigh type Surface Acoustic Waves. Using the
Transient Grating geometry\cite{Rogers, Janusonis, Koralek,
Tobey}, we were able to demonstrate frequency tunability from
1~GHz to $\approx$6~GHz. Here we demonstrate the broader utility
of the TG technique for generating  additional planar elastic
waves, beyond Rayleigh SAW, which also drive magnetization
precession.  The TG geometry generates all elastic modes that
satisfy the boundary condition imposed by the thermoelastic
stress, regardless of frequency. Therefore, in our geometry we
access additional elastic excitations known as Surface Skimming
Longitudinal Waves. Under such excitation conditions, both elastic
waves are excited simultaneously, and we show the magnetic field
selective coupling of the magnetization precession to each elastic
wave independently.

In the TG geometry, short pulses of light at 400~nm are crossed
onto the sample surface, which upon superposition result in a
spatially periodic excitation of the sample, as shown in
Fig.~\ref{fig1}.  In this excitation geometry, all elastic modes
that satisfy the elastic boundary conditions are excited at the
wavevector $2\pi/\Lambda$ determined by the crossing of two beams.
Subsequent to the excitation, a time-delayed probe pulse impinges
normally onto the sample surface and the transmitted radiation is
polarization analyzed (Faraday detection). The sample in held in a
magnetic field that can be swept continuously from -1.5~kG to
+1.5~kG and rotated around the sample normal.

\begin{figure}
\includegraphics[width=\columnwidth]{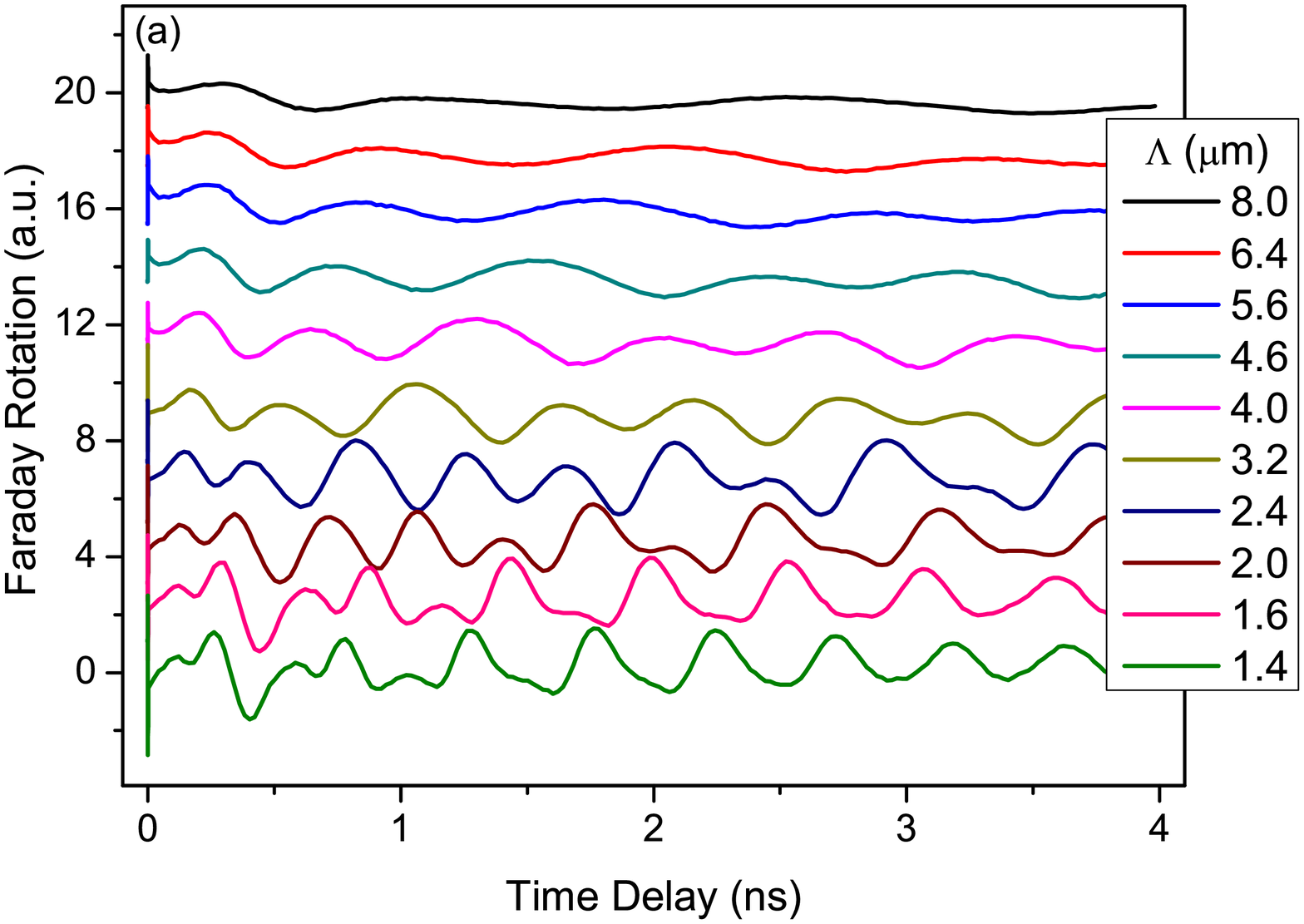}
\includegraphics[width=\columnwidth]{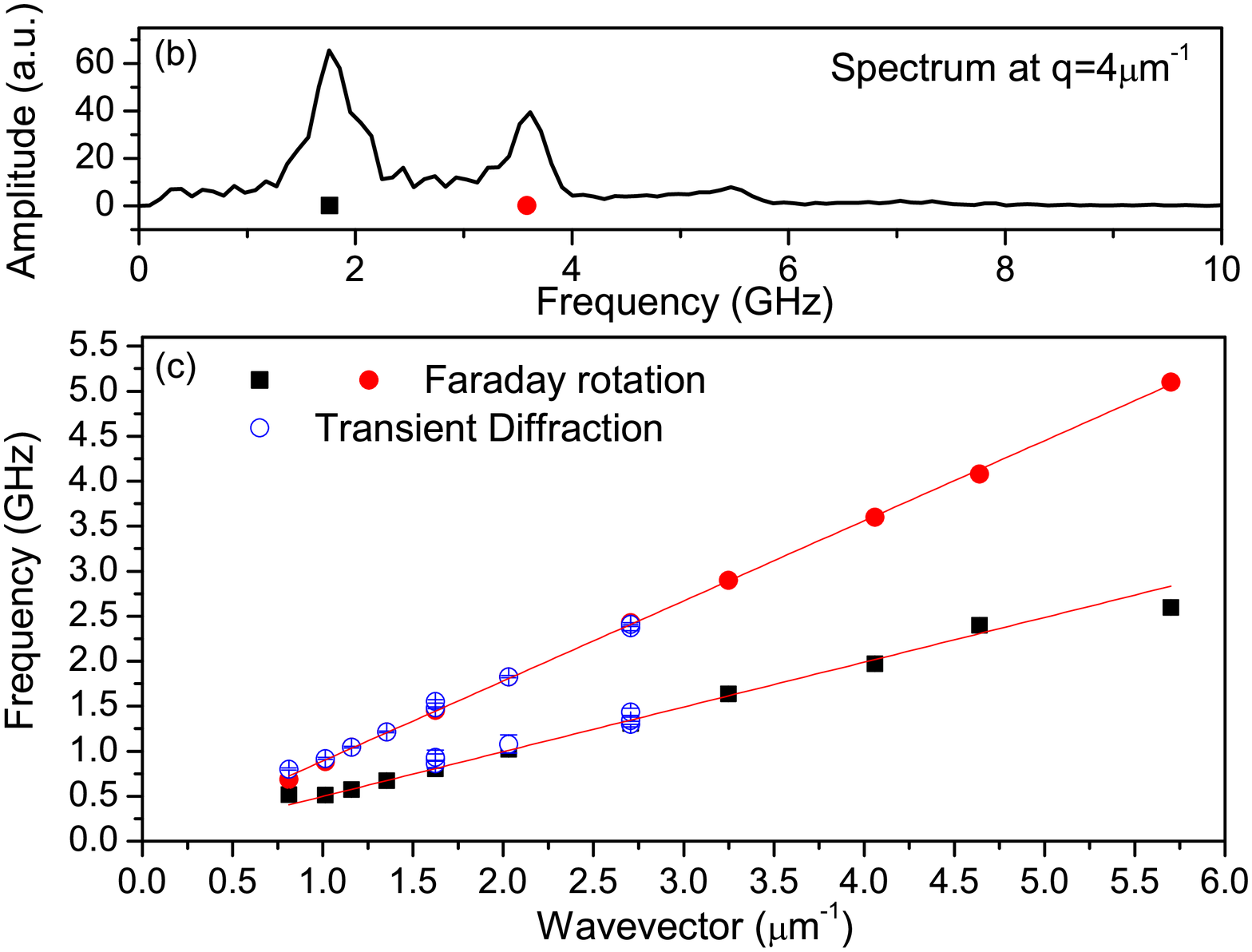}
\caption{ (a) The time-resolved Faraday rotation measures the
acoustically induced magnetization precession. As the period of
the transient grating is increased, the precession decreases in
frequency and amplitude. In the limit of a single pump beam
excitation, the magnetic precession is completely suppressed. (c)
The frequency of oscillation versus wavevector measures the
velocity of acoustic propagation.  Red and black data points
represent frequencies extracted from (a), and blue data points are
extracted from the nonmagnetic transient grating detection. Near
perfect linear dependence allows us to identify two types of
excitations as the Rayleigh Surface Acoustic Wave (lower branch)
and the Surface Skimming Longitudinal Wave (upper branch)
propagating at $3120 \pm 20$~m/s and $5590 \pm 15$~m/s,
respectively.} \label{fig2}
\end{figure}

The samples under study are composed of thin polycrystalline
nickel films on glass substrates (soda lime glass). Representative
data showing the grating dependence of the Faraday response are
shown in Fig.~\ref{fig2}(a) for a film thickness of 60nm.  In
contrast to our previous results\cite{Janusonis} which were
reported for the nickel/MgO sample structure, when performing
measurements on glass substrates additional dynamics can be
observed. Cursory evaluation of the data in Fig.~\ref{fig2}(a)
shows that both oscillation amplitude and frequency reduce as the
excitation grating period is increased, and eventually disappear
when a single pump beam is used to excite the material.  Secondly,
the dynamics in the first nanosecond are comprised of a two
oscillating contributions, which suggests the presence of two
distinct magnetoelastic waves.

Assignment of the modes is accomplished by plotting the
frequencies of both modes as a function of excitation wavevector.
In Fig.~\ref{fig2}(b) we display the Fourier Transform of the zero
field response showing two oscillations. For $\Lambda =
1.57~\mu$m, and correspondingly $k = 4~\mu{\rm m}^{-1}$, the
extracted frequencies are 1.75~GHz and 3.6~GHz, respectively. In
Fig.~\ref{fig2}(c) these frequencies are plotted for a range of
applied grating periodicities.   Overlayed onto the Faraday
response are data acquired in the conventional transient grating
geometry, where light is diffracted from acoustic waves (data not
shown, see Ref.5 for details). Such an experimental scheme is
known to be sensitive to the underlying elastic waves and their
associated structural distortions. From the correspondence between
the two detection schemes, we are able to fit a single linear
relationship (with zero intercept) providing the following mode
assignments and their respective velocities: The lower branch
propagates at $3120 \pm$20~m/s which we assign as the Rayleigh
Surface Acoustic Wave, the same excitation witnessed on MgO
substrates\cite{Janusonis}.  In the long wavelength limit
($\Lambda > 1~\mu$m), the propagation velocity of the Rayleigh SAW
in the Ni/substrate heterostructure is dictated by the substrate
elastic constants due to the finite penetration depth
($\approx\Lambda\gg d_{Ni}$) of the elastic wave.  This velocity
compares favorably with the Rayleigh SAW velocity of the glass
substrates, 3100~m/s for soda lime glass.   The upper branch,
propagating at a velocity of $5590 \pm 15$~m/s, is the in-plane
longitudinal acoustic wave, which has been termed previously as a
Surface Skimming Longitudinal Wave (SSLW)\cite{Xu}, or the Surface
Skimming Bulk Wave (SSBW)\cite{Lewis}, an elastic excitation that
has been used extensively for non-destructive material
evaluation\cite{Sathish, Abbas}.   Again, the velocity is very
close to the longitudinal sound velocity in glass (literature
value 5400~m/s) due to the predominant concentration of elastic
energy in the substrate.

\begin{figure}
\includegraphics[width=\columnwidth]{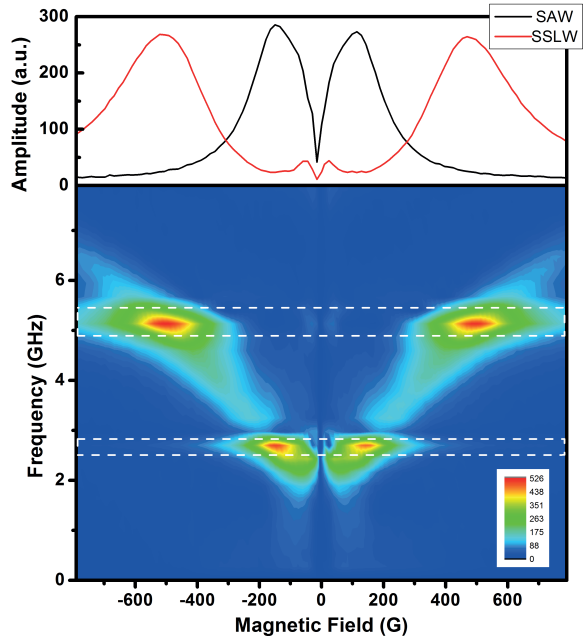}
\caption{Under appropriate field conditions the time-periodic
effective magnetic field of the passing elastic wave resonantly
couples to the precessional motion in the film, driving large
amplitude precessional motion. For $\Lambda = 1.1~\mu$m, the upper
frequency at 5.1~GHz is that of the SSLW while the lower frequency
at 2.8~GHz corresponds to the SAW. In the upper panel,
vertically-averaged signal within the indicated boxes for both SAW
and SSLW is displayed.} \label{fig3}
\end{figure}

\begin{figure}
\includegraphics[width=1\columnwidth]{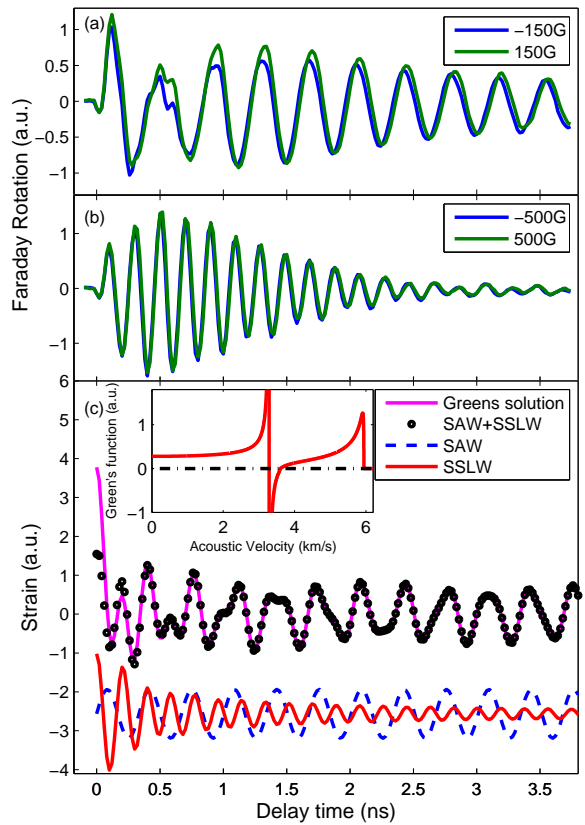}
\caption{ Time traces at the resonances for (a) SAW and (b) SSLW,
taken from Fig.~\ref{fig3}(main panel).  Switching the polarity of
the magnetic field yields nearly identical oscillation amplitudes
and phases which can be understood by considering the applied
torque induced by magnetoelastic coupling. (c) The time-dependent
strain $\epsilon_{xx}(t)$ calculated by Green's function formulism
(discussed in the Supplemental Material) confirms the simultaneous
excitation of SAW and SSLW modes. Whereas the decay of the
acoustic eigenmode SAW is negligible, the amplitude of SLLW
decreases as $1/t$, as expected for an acoustic diffraction
phenomenon in a cylindric geometry.} \label{fig4}
\end{figure}

We now discuss the effect of applying an in-plane magnetic field.
As we had shown previously\cite{Janusonis}, applying a magnetic
field $\overrightarrow{H}_{ext}$ in the plane of the sample
provides the coupling between the elastic field at frequency
$v_{ac}/\Lambda$ and the ferromagnetic resonance of the film at
frequency $f_{\rm FMR}\propto\sqrt{H_{ext}(H_{ext}+M)}$.  The
magnitude $H_{ext}$ of the applied field tunes the precessional
FMR frequency of the film to match either SAW or SSLW frequency
($M$ is the saturation magnetization of the thin magnetic film).
The effect of having two elastic fields present, provides for the
selective excitation of one magnetization precession response.  A
representative field scan is shown in Fig.~\ref{fig3}(main panel,
the Fourier Transform amplitude of the Faraday signal) for a
grating period of 1.1~$\mu$m for both polarities of the applied
field. The elastic frequencies at $\Lambda$ =  $1.1~\mu$m
associated with SAW and SSLW are indicated by rectangles over the
data, while lineouts of the vertically integrated values in the
bounding boxes are displayed in Fig.~\ref{fig3}(top). As the
applied field is tuned into resonance, the oscillation amplitude
of the magnetization precession peaks, and then reduces as the
field is tuned above resonance.  For all applied fields, it should
be recognized that two elastic waves are active, but the resonance
condition drives a single precessional motion of the
magnetization.  The maxima occur at the location where the elastic
driving frequency matches FMR frequency.

The temporal responses at the two resonant conditions, $\pm$150~G
and $\pm$500~G, are shown in Fig.~\ref{fig4}, exhibiting the same
amplitude and phase of oscillation for both polarities of the
magnetic field.  The correspondence in oscillation phase for two
directions of the magnetic field can be understood by considering
the torque applied to the magnetization via the magnetoelastic
coupling, through the Landau-Lifshits-Gilbert equation: $\partial
\overrightarrow{M} / \partial t \propto \overrightarrow{M} \times
[\overrightarrow{H}_{ext}+\overrightarrow{H}_{shape}+\overrightarrow{H}_{me}(t)]$,
where the time-periodic effective field
$\overrightarrow{H}_{me}(t)$ is determined by magnetoelastic
interactions and can be calculated as a derivative of a the
magneto-elastic term in the free energy density with respect to
magnetization \cite{Kovalenko, Dreher}. The out-of-plane torque
component originates from the coupling of in-plane longitudinal
strain ($\epsilon_{xx}$(t), the primary strain component in the
SSLW and the dominant component in the SAW wave) and the in-plane
components $M_x$ and $M_y$ of magnetization vector: $\partial
\overrightarrow{M}_z / \partial t \propto
\epsilon_{xx}(t)M_xM_y$. When the magnetization is inverted, i.e.
$M_x\rightarrow-M_x$ and $M_y\rightarrow-M_y$, the magnetoelastic
torque direction, and therefore the precessional direction,
remains unchanged. On the contrary, if the magnetization is
reflected with respect the acoustic wavevector ($M_x \rightarrow
M_x$ and $M_y\rightarrow-M_y$, which is equivalent to
$\phi\rightarrow-\phi$) the direction of magnetization precession
changes in accordance with the above equation (Fig.~1S in the
Supplemental Material). We note that the foregoing discussion on
magnetoelasticity is related to a myriad of other ultrafast
work\cite{Scherbakov, Linnik, Bombeck, Kim, Jaeger1,Kovalenko,
Jaeger2} as well as it's connection to quasi-static
'straintronics'\cite{Roy, Thevenard}.

The differences in lifetime of the respective modes
(Fig.~\ref{fig4})(a),(b) can be traced directly to the nature of
the elastic driving field underscoring these effects.  Whereas the
SAW is a surface propagating elastic eigenwave with negligible
energy dissipation, the SSLW is not a surface bound wave in this
strict sense, and thus significant elastic energy propagates away
from the active magnetic layer. Thus the strain amplitude of this
higher frequency mode decays rapidly as elastic energy leaks away
from the magnetically active material.

Qualitatively similar results can be found by solving the Green's
Function response of an elastic half space, which at present we
consider to be free of the bounding magnetic film\cite{Lee}.
Convolution of the impulse response with spatially periodic stress
associated with $\Lambda$ results in a series of elastic
excitations, most notably the SAW and SSLW that reside along the
surface of the sample.  The results of such a calculation
(discussed in more detail in the Supplemental Material) are
displayed in figure Fig.~\ref{fig4}(c).  The time evolution for
the strain amplitude can be robustly reproduced by assuming
independent elastic waves (SAW and SSLW), while there sum shows
excellent correspondence with the Green's function result.  The
two elastic constituents are assumed to be the negligibly decaying
SAW and highly damped SSLW, which decays with 1/time dependency in
accordance with the cylindrical symmetry of the grating
excitation.  Thus, both experiment and theory corroborate the
coexistence of elastic waves, while field tuning allows us to
monitor each wave independently through it's magnetoelastic
coupling.

Finally, we note that measurements performed in similar planar
geometries using elastic transducers to generate elastic waves and
drive magnetic precession\cite{Davis, Thevenard2, Weiler2, Uchida}
do not witness similar SSLW excitations. In the TG geometry, we
launch all elastic modes which satisfy the boundary conditions
imposed by the spatially periodic stress conditions.  The
transducer measurements additionally control the driving
frequency, and thus primarily excite the SAW excitation only.
Furthermore, as evident from our time traces, the SSLW excitations
at multi-GHz frequencies experience far larger decay rates than
SAW and thus would inhibit their detection in a non-local geometry
based on generation and detection transducers that are spatially
separated.  In our local measurements, where the excitation and
detection are performed in the same position, enhance the
detection of such short lived elastic waves.

In summary, we have demonstrated the generation of two distinct
elastic waves using a single excitation geometry, based on the
transient grating technique. The two modes are identified as the
surface-bound Rayleigh Surface Acoustic Wave (SAW) and the leaky
Surface Skimming Longitudinal Wave (SSLW).  Both elastic
excitations couple to and drive the magnetization precession in a
resonant fashion when an appropriate magnetic field is applied.
Furthermore, since both elastic distortions are active
simultaneously,  we demonstrated the field tuned selectivity of
each magnetoelastic excitation.  The current measurements are
distinguished by their time-domain approach which allowed us to
witness the coupling between acoustic and magnetic degrees of
freedom in real time.  We envisage further experimental efforts to
focus on extraordinary transmission of acoustic waves through
subwavelength apertures while the magnetoelastic detection scheme
provides for individual sensitivity to both elastic wave effects,
opening possibilities to study competition between different
elastic contributions to transmission measurements.

Authors thank K. A. Nelson, A.A. Maznev, and J.Y. Duquesne for
discussions.  We thank M. de Roosz for assistance with e-beam
sample preparation.  Funding from Nouvelle \'{e}quipe, nouvelle
th\'{e}matique et Strat\'{e}gie internationale NNN-Telecom de la
R\'{e}gion Pays de La Loire, Russian Foundation for Basic Research
(Grant No. 15-02-07575) and Alexander von Humboldt Stiftung is
gratefully acknowledged.

\bibliography{Janusonis}

\end{document}